\documentclass[aps,reprint,groupedaddress,nofootinbib]{revtex4-1}
\pdfoutput=1
\usepackage{amsmath}
\usepackage{amssymb}
\usepackage{slashed}
\usepackage[colorlinks=true,linkcolor=red]{hyperref}
\usepackage{cleveref}
\usepackage{eso-pic}

\newcommand{\cA}{\mathcal{A}}
\newcommand{\cL}{\mathcal{L}}
\newcommand{\cO}{\mathcal{O}}

\begin{document}


\title{The double copy for heavy particles}


\author{Kays Haddad}
\email[]{kays.haddad@nbi.ku.dk}
\author{Andreas Helset}
\email[]{ahelset@nbi.ku.dk}
\affiliation{Niels Bohr International Academy \& Discovery Center, Niels Bohr Institute, Univesity of Copenhagen, Blegdamsvej 17, DK-2100, Copenhagen, Denmark}


\date{\today}

\begin{abstract}
	We show how to double-copy Heavy Quark Effective Theory (HQET) to Heavy Black Hole Effective Theory (HBET) for spin $s\leq 1$.
	In particular, the double copy of spin-$s$ HQET with scalar QCD produces spin-$s$ HBET, 
	while the double copy of spin-1/2 HQET with itself gives spin-1 HBET.
	Finally, we present novel all-order-in-mass Lagrangians for spin-1 heavy particles.
\end{abstract}


\maketitle

\AddToShipoutPictureBG*{%
  \AtPageUpperLeft{%
    \hspace{\paperwidth}%
    \raisebox{-1.75\baselineskip}{%
      \makebox[0pt][r]{SAGEX-20-13-E\qquad}
}}}%

\section{Introduction}

An expanding family of field theories has been observed to obey double-copy\footnote{For a review of the double-copy program, see ref.~\cite{Bern:2019prr}.} relations \cite{Kawai:1985xq,Bern:2008qj,Bern:2010ue,Bern:2010yg,Bern:1999bx,Broedel:2012rc,Chen:2013fya,Johansson:2014zca,Chiodaroli:2014xia,Cachazo:2014xea,Chen:2014dfa,Du:2016tbc,Chiodaroli:2015rdg,Casali:2015vta,Stieberger:2016lng,Schlotterer:2016cxa,Nandan:2016pya,Cheung:2017yef,Bergshoeff:1996tu,Bergshoeff:1997kr,Kallosh:1997aw,Bergshoeff:2013pia,Cachazo:2016njl,He:2016mzd,Elvang:2018dco,Cheung:2016prv,Chiodaroli:2017ngp,Chiodaroli:2017ehv,Teng:2017tbo,Du:2017gnh,Cheung:2017ems,Chiodaroli:2018dbu}.
In particular, scattering amplitudes of gravitational theories with massive matter can be calculated
from the double copy of gauge theories with massive matter \cite{Johansson:2015oia,delaCruz:2015dpa,Luna:2017dtq,Brown:2018wss,Plefka:2018zwm,Bern:2019nnu,Bern:2019crd,Bautista:2019tdr,Johansson:2019dnu,Bautista:2019evw,Plefka:2019wyg}.

As Heavy Quark Effective Theory (HQET) \cite{Georgi:1990um} is derived from QCD and Heavy Black Hole Effective Theory (HBET) \cite{Damgaard:2019lfh} is derived from gravity coupled to massive particles, the amplitudes of HBET should be obtainable as double-copies of HQET amplitudes.
Indeed, this is the main result of this paper.
We show through direct computation that the three-point and Compton amplitudes
of HQET and HBET satisfy the schematic relations
\begin{subequations}\label{eq:DCResults}
\begin{align}
    \label{eq:0xsDCResult}
    (\text{QCD}_{s=0})\times(\text{HQET}_{s})&=\text{HBET}_{s}, \\
    \label{eq:1/2x1/2DCResult}
    (\text{HQET}_{s=1/2})\times(\text{HQET}_{s=1/2})&=\text{HBET}_{s=1},
\end{align}
\end{subequations}
for $s\leq 1$, where the spin-$s$ HQET and HBET matter states are equal in the free-field limit,
and the spin-1 heavy polarization vectors are related to the heavy spinors through \cref{eq:PolVecDef}.
While we only show here the double copy for three-point and Compton amplitudes,
invariance of the $S$-matrix under field redefinitions implies that
\cref{eq:DCResults} holds more generally whenever QCD double-copies to gravitationally interacting matter.
\Cref{eq:DCResults} expands the double copy in powers of $\hbar$
since the operator expansion for heavy particles can be interpreted as an expansion in $\hbar$ \cite{Damgaard:2019lfh}.
The $\hbar\rightarrow0$ limit of the double copy is currently of particular relevance \cite{Luna:2017dtq,Bern:2019nnu,Bern:2019crd}.

We will begin in \Cref{sec:CKRev} with a brief review of the color-kinematics duality,
and we will also discuss double-copying with effective matter fields. In \Cref{sec:Spin0Amps,sec:Spin1/2Amps,sec:Spin1Amps} we demonstrate the double copy at tree level for three-point and Compton amplitudes for spins 0, 1/2, and 1, respectively. We conclude in \Cref{sec:Conclusion}. The Lagrangians used to produce the amplitudes in this paper are presented in \Cref{sec:Lagrangians}.
Among them are novel all-order-in-mass Lagrangians for spin-1 HQET and HBET given in \cref{eq:LagrGluons1,eq:LagrGravitons1}.

\section{Color-kinematics duality and heavy fields}
\label{sec:CKRev}

An $n$-point gauge-theory amplitude, potentially with external matter, can be written as\footnote{
We omit coupling constants for the sake of clarity. Reinstating them is straight-forward: after double-copying the gauge theory coupling undergoes the replacement $g\rightarrow\sqrt{\kappa}/2$.}
\begin{align}\label{eq:CKYM}
    \mathcal{A}_{n}=\sum_{i\in\Gamma}\frac{c_{i}n_{i}}{d_{i}},
\end{align}
where $\Gamma$ is the set of all diagrams with only cubic vertices.
Also, $c_{i}$ are color factors, $n_{i}$ encode the kinematic information, and $d_{i}$ are propagator denominators.
A subset of the color factors satisfies the identity
\begin{align}
    c_{i} + c_{j} + c_{k}=0.
\end{align}
If the corresponding kinematic factors satisfy the analogous identity,
\begin{align}
	\label{eq:dualck}
    n_{i} + n_{j} + n_{k}=0,
\end{align}
and have the same anti-symmetry properties as the color factors,
then the color and kinematic factors are dual. In this case,
the color factors in \cref{eq:CKYM} can be replaced by kinematic
factors to form the amplitude
\begin{align}\label{eq:CKGR}
	\mathcal{M}_{n}=\sum_{i\in\Gamma}\frac{n^{\prime}_{i}n_{i}}{d_{i}},
\end{align}
which is a gravity amplitude with anti-symmetric tensor and dilaton contamination.\footnote{For an amplitude of arbitrary multiplicity containing massive external states with an arbitrary spectrum, \cref{eq:CKGR} may not represent a physical amplitude \cite{Johnson:2020pny}.
However, for the cases under consideration in this paper, the application of the double copy will yield a well-defined gravitational amplitude.}
In general, $n^{\prime}_{i}$ and $n_{i}$ need not come from the same gauge theory,
and only one of the sets must satisfy the color-kinematics duality.


In this paper we are interested in applying the double-copy procedure to HQET.
A complicating factor to double-copying effective field theories (EFTs) is that 
Lagrangian descriptions of EFTs are not unique,
as the Lagrangian can be altered by redefining one or more of the fields.
The LSZ procedure \cite{Lehmann:1954rq} guarantees the invariance of the $S$-matrix, and in particular \cref{eq:CKYM,eq:CKGR}, 
under such field redefinitions by accounting for 
wavefunction normalization factors (WNFs) $\mathcal{R}^{-1/2}$,
which contribute to the on-shell residues of two-point functions.\footnote{Note that $\mathcal{R}^{-1/2}=1$ for canonically normalized fields. The WNF for an effective state $\tilde{\varepsilon}$ can thus be determined by relating it to a canonically normalized state $\varepsilon$ through
\begin{align}
	\label{eq:MatchingCondition}
	\varepsilon = \mathcal{R}^{-1/2}\cdot \tilde\varepsilon.
\end{align}}
Under the double copy the WNFs from each matter copy
combine in a spin-dependent manner, which complicates the matching
of the double-copied amplitude to one derived from a gravitational Lagrangian.

In order to ease the double-copying of HQET to HBET,
we would like to avoid having to compensate for the WNFs.
This can be achieved by ensuring that HQET and HBET have the same 
WNFs -- i.e. that the asymptotic states
for the spin-$s$ particles in HQET and HBET are equal --
and double-copying HQET with QCD, which has a trivial WNF.

The asymptotic states -- that is, the states in the free-field limit -- of the canonically normalized theories (given by complex Klein-Gordon, Dirac, and symmetry-broken Proca actions) are related to their respective asymptotic heavy states (labelled by a velocity $v$) in position-space through
\begin{subequations}
	\label{eq:AsympFields}
\begin{align}
    \label{eq:AsympScalars}
	\varphi(x)&= \frac{e^{-imv\cdot x}}{\sqrt{2m}} \left[1-\frac{1}{2m +iv\cdot \partial + 
	\frac{\partial^2_\perp}{2m}} \frac{\partial_\perp^2}{2m}\right]\phi_{v}(x),	 \\
	\label{eq:AsympSpinors}
    \psi(x)&=e^{-imv\cdot x}\left[1+\frac{i}{2m+iv\cdot\partial}(\slashed{\partial}-v\cdot\partial)\right]Q_{v}(x),\\
    \label{eq:AsympVectors}
    A^\mu(x) &=\frac{e^{-imv\cdot x}}{\sqrt{2m}} \left[ \delta^\mu_\nu - \frac{iv^\mu \partial_\nu-\partial^{\mu}\partial_{\nu}/2m}{m+iv\cdot \partial/2}\right]B^\nu_{v}(x),
\end{align}
\end{subequations}
where $a_{\perp}^{\mu}=a^{\mu}-v^{\mu}(v\cdot a)$ for a vector $a^\mu$. 
Here, the momentum is decomposed as $p^\mu = m v^\mu + k^\mu$ in the usual heavy-particle fashion.
The Lagrangians for the heavy fields in \cref{eq:AsympFields} are given in \Cref{sec:Lagrangians}.
Converting to momentum space, 
\cref{eq:AsympFields} gives the WNFs 
\begin{subequations}
\begin{align}
    \label{eq:InvR0}
	\mathcal{R}_{s=0}^{-1/2}(p) &=\frac{1}{\sqrt{2m}}\left[1+\frac{k_\perp^2}{4m^{2}+2mv\cdot k - k_\perp^2}\right], \\
	\label{eq:InvR1/2}
	\mathcal{R}_{s=1/2}^{-1/2}(p) &= 1+\frac{1}{2m+v\cdot k}(\slashed k-v\cdot k), \\
	\label{eq:InvR1}
	\left(\mathcal{R}_{s=1}^{-1/2}(p)\right)^{\nu}_{\mu} &=\frac{1}{\sqrt{2m}}\left[ \delta^{\nu}_\mu-\frac{v_{\mu}k^{\nu}+k_{\mu}k^{\nu}/2m}{m+v\cdot k/2}\right].
\end{align}
\end{subequations}
%
%

We will demonstrate that spin-$s$ HBET amplitudes can directly be obtained by double-copying spin-$s$ HQET amplitudes with scalar QCD 
for spins $s\leq 1$.
At $s=1$ there is also the possibility to double-copy using two spin-1/2 amplitudes.
We will discuss this point further below.

\section{Spin-0 Gravitational Amplitudes}
\label{sec:Spin0Amps}

We begin with the simplest case of spinless amplitudes.
%
Consider first the three-point amplitude. For scalar HQET we have that
\begin{align}
	\label{eq:3ptsGluonHScalar}
	\cA_3^{\text{H},s=0} =& - \mathbf{T}^a_{ij} \epsilon_q^{*\mu} \phi_{v}^*
\left(1+\frac{k_1^2+k_2^2}{4m^2}\right)\phi_{v} \nonumber \\
	&\times	\left[
		v_\mu + \frac{\left(k_1+k_2\right)_\mu}{2m}
	\right]
	+ \cO\left(m^{-4}\right),
\end{align}
where $k_{2}=k_{1}-q$. For scalar QCD the amplitude is
\begin{align}
    \label{eq:3ptsGluonScalar}
	\cA_3^{s=0} =& - \mathbf{T}^a_{ij} \epsilon_q^{*\mu} \left[
		2m v_\mu + \left(k_1+k_2\right)_\mu
	\right].
\end{align}
Note that we have left the external heavy scalar factor $\phi_v$ explicit in the HQET amplitude.
This is because, in contrast to the canonically normalized scalar fields, the heavy scalar factors are not equal to $1$ in momentum space.
Indeed, for the HQET amplitude to be equal to the QCD amplitude, the heavy scalar factor in momentum space must be equal to the inverse of \cref{eq:InvR0}.
This will cancel the extra factor in round brackets in \cref{eq:3ptsGluonHScalar}.

The double copy at three-points is simply given by a product of amplitudes:
\begin{align}
    \label{eq:3ptsGluonDCScalar}
    &\mathcal{A}_{3}^{s=0}\mathcal{A}_{3}^{\text{H},s=0}=\epsilon_{q}^{*\mu}\epsilon_{q}^{*\nu}\phi^{*}_{v}\left(1+\frac{k_1^2+k_2^2}{4m^2}\right)\phi_{v}\notag \\
    &\ \times 2m\left[v_{\mu}v_{\nu}+v_{\mu}\frac{k_{1\nu}+k_{2\nu}}{m}+\frac{(k_{1}+k_{2})_{\mu}(k_{1}+k_{2})_{\nu}}{4m^{2}}\right]\notag \\
    &\ +\mathcal{O}(m^{-3}).
\end{align}
As the only massless particle in this process is external, we can easily eliminate the massless non-graviton degrees of freedom by identifying the outer product of gluon polarization vectors with the graviton polarization tensor. 
After doing so, \cref{eq:3ptsGluonDCScalar} agrees with the three-point amplitude derived from \cref{eq:LagrGravitons0}.

As another example, consider the Compton amplitude.
The color decomposition for Compton scattering\footnote{We have computed all Compton amplitudes using NRQCD propagators.
It is also possible to perform the computations using HQET propagators: in that case, a comparison to the Compton amplitude for the emission of bi-adjoint scalars from heavy particles (described by the Lagrangians in \cref{eq:LagrBiadjoints0,eq:LagrBiadjoints1/2,eq:LagrBiadjoints1}) -- analogous to the treatment in ref.~\cite{Shen:2018ebu} -- is necessary to identify kinematic numerators. Both methods produce the same results. }
%
is
\begin{subequations}
\begin{align}
	\label{eq:colorDecomp}
	\cA_4^s = \frac{c_s n_s}{d_s} + \frac{c_t n_t}{d_t} + \frac{c_u n_u}{d_u},
\end{align}
where
\begin{align}
	\label{eq:colorStruct}
	c_s = {\bf T}_{ik}^a {\bf T}_{kj}^b,  \quad
	c_t = i f^{abc} {\bf T}_{ij}^c, \quad
	c_u = {\bf T}_{ik}^b {\bf T}_{kj}^a.
\end{align}
\end{subequations}
The kinematic numerators for scalar HQET are
\begin{subequations}
\begin{align}
	\label{eq:4ptsGluonScalar}
	n_s^{\text{H},s=0} &= -2m \phi_v^* \epsilon_{q_1}^{*\mu} \epsilon_{q_2}^{*\nu}
		v_\mu v_\nu 
	\left(1+\frac{k_1^2+k_2^2}{4m^2}\right)\phi_v, \\
	n_t^{\text{H},s=0} &=0, \\
	n_u^{\text{H},s=0} &= n_s^{\text{H},s=0}|_{q_1\leftrightarrow q_2},
\end{align}
\end{subequations}
where $k_{2}=k_{1}-q_{1}-q_{2}$. Those for scalar QCD are
\begin{subequations}
\begin{align}
    n_{s}^{s=0}&=-4m^2 \epsilon_{q_1}^{*\mu} \epsilon_{q_2}^{*\nu} 
		v_\mu v_\nu, \\
    n_{t}^{s=0}&=0, \\
    n_{u}^{s=0}&=n_s^{s=0}|_{q_1\leftrightarrow q_2}.
\end{align}
\end{subequations}
For brevity we have written the numerators under the conditions $k_{1}=q_{i}\cdot\epsilon_{j}=\epsilon_{i}\cdot\epsilon_{j}=0$;
the initial residual momentum can always be set to 0 by reparameterizing $v$,
and such a gauge exists for opposite helicity gluons.
We have checked explicitly up to and including $\mathcal{O}(m^{-2})$
that the following results hold when relaxing all of these conditions.

Both the HQET and QCD numerators satisfy the color-kinematics duality in the form
\begin{align}\label{eq:ComptCKDuality}
    c_{s}-c_{u}=c_{t}\ \Leftrightarrow\ n_{s}-n_{u}=n_{t}.
\end{align}
We can therefore replace the color factors in the HQET amplitude with the QCD kinematic numerators,
\begin{align}
    \label{eq:ComptDCScalar}
    \mathcal{M}_{4}^{\text{H},s=0}&=\frac{n_{s}^{s=0}n_{s}^{\text{H},s=0}}{d_{s}}+\frac{n_{t}^{s=0}n_{t}^{\text{H},s=0}}{d_{t}}+\frac{n_{u}^{s=0}n_{u}^{\text{H},s=0}}{d_{u}}.
\end{align}
Identifying once again the outer products of gluon polarization vectors with graviton polarization tensors,
we find that the Compton amplitude derived from \cref{eq:LagrGravitons0} agrees with \cref{eq:ComptDCScalar}.

To summarize, we have explicitly verified that
\begin{align}
    (\text{QCD}_{s=0})\times(\text{HQET}_{s=0})=\text{HBET}_{s=0}
\end{align}
for three-point and Compton amplitudes.

\section{Spin-1/2 Gravitational Amplitudes}
\label{sec:Spin1/2Amps}

We now move on to the double copy of spin-1/2 HQET with scalar QCD to obtain spin-1/2 HBET.
The three-point spin-1/2 HQET amplitude is
\begin{align}
    \label{eq:3ptsGluonSpinor}
    \mathcal{A}^{\text{H},s=\frac{1}{2}}_{3}&=-\mathbf{T}^{a}_{ij}\bar{u}_{v}u_{v}\epsilon_{q}^{*\mu}\left(v_{\mu}+\frac{k_{1\mu}}{m}+\frac{k_1^{2}-k_1\cdot q}{4m^{2}}v_{\mu}\right)\notag \\
    &\quad-\frac{i\mathbf{T}^{a}_{ij}}{2m}\bar{u}_{v}\sigma^{\alpha\beta}u_{v}\epsilon_{q}^{*\mu}\left[q_{\alpha}\eta_{\beta\mu}-\frac{1}{2m}q_{\alpha}k_{1\beta}v_{\mu}\right]
    \notag \\ &\quad + \mathcal{O}(m^{-3})
\end{align}
Double-copying with scalar QCD, we find
\begin{align}
    \mathcal{M}^{\text{H},s=\frac{1}{2}}_{3}&=\mathcal{A}^{s=0}_{3}\mathcal{A}^{\text{H},s=\frac{1}{2}}_{3},
\end{align}
where $\mathcal{M}^{\text{H},s=\frac{1}{2}}_{3}$ is the amplitude derived from \cref{eq:LagrGravitons1/2}.

We turn now to Compton scattering. For brevity we write here the amplitudes in the case $k_{1}=q_{i}\cdot\epsilon_{j}=\epsilon_{i}\cdot\epsilon_{j}=0$.
We have checked explicitly that the results hold when these conditions are relaxed.
Also, we have performed the calculation up to $\mathcal{O}(m^{-2})$ but only present the kinematic numerators up to $\mathcal{O}(m^{-1})$.
They are
\begin{subequations}
\begin{align}
    n^{\text{H},s=\frac{1}{2}}_{s}&=-2m\bar{u}_{v}\left[v\cdot\epsilon_{q_{1}}^{*}v\cdot\epsilon_{q_{2}}^{*}\right.\\
    &\ \left.-\frac{iv_{\rho}}{2m}\sigma_{\mu\nu}(\epsilon_{q_{1}}^{*\mu}q_{1}^{\nu}\epsilon_{q_{2}}^{*\rho}+\epsilon_{q_{2}}^{*\mu}q_{2}^{\nu}\epsilon_{q_{1}}^{*\rho}-q_{2}^{\rho}\epsilon_{q_{2}}^{*\mu}\epsilon_{q_{1}}^{*\nu})\right]u_{v}, \notag\\
    n^{\text{H},s=\frac{1}{2}}_{t}&=0, \\
    n^{\text{H},s=\frac{1}{2}}_{u}&=n^{\text{H},s=\frac{1}{2}}_{s}|_{q_{1}\leftrightarrow q_{2}}.
\end{align}
\end{subequations}
In this case, the color-kinematic duality \cref{eq:ComptCKDuality} is violated at $\mathcal{O}(m^{-2})$.
Nevertheless, since the scalar QCD kinematic numerators satisfy the duality we can use them to double copy the spin-1/2 Compton amplitude. 
Doing so we find
\begin{align}
    \mathcal{M}^{\text{H},s=\frac{1}{2}}_{4}&=\frac{n_{s}^{s=0}n_{s}^{\text{H},s=\frac{1}{2}}}{d_{s}}+\frac{n_{t}^{s=0}n_{t}^{\text{H},s=\frac{1}{2}}}{d_{t}}+\frac{n_{u}^{s=0}n_{u}^{\text{H},s=\frac{1}{2}}}{d_{u}},
\end{align}
where $\mathcal{M}^{\text{H},s=\frac{1}{2}}_{4}$ is the spin-1/2 HBET Compton amplitude derived from \cref{eq:LagrGravitons1/2}.

We have seen that
\begin{align}
    (\text{QCD}_{s=0})\times(\text{HQET}_{s=1/2})=\text{HBET}_{s=1/2}
\end{align}
for the three-point and Compton amplitudes.

\section{Spin-1 Gravitational Amplitudes}
\label{sec:Spin1Amps}

Gravitational amplitudes with spin-1 matter can be obtained by double-copying two gauge theories
with matter in two ways: spin-0 $\times$ spin-1 or spin-1/2 $\times$ spin-1/2 \cite{Johansson:2019dnu,Bautista:2019evw,Bautista:2019tdr}.
This fact also holds for heavy particles.
We now show this in two examples by deriving the spin-1 gravitational three-point and Compton amplitudes
using both double-copy procedures.

\subsection{\texorpdfstring{$0\times1$}{0x1} Double Copy}

The three-point spin-1 HQET amplitude is
\begin{align}
    \mathcal{A}^{\text{H},s=1}_{3}&=\mathbf{T}^{a}_{ij}\varepsilon_{v}^{*\beta}\varepsilon_{v}^{\alpha}\epsilon_{q}^{*\mu}\left[\eta_{\alpha\beta}v_{\mu}\right.\notag \\
				  &\quad+\frac{1}{2m}(\eta_{\alpha\beta}(k_1+k_2)_{\mu}-2q_{\beta}\eta_{\alpha\mu}+2q_{\alpha}\eta_{\beta\mu})\notag \\
    &\quad\left.+\frac{1}{2m^{2}}v_{\mu}(-k_{1\beta}q_{\alpha}+q_{\alpha}q_{\beta}+q_{\beta}k_{1\alpha})\right],
\end{align}
where $k_{2}^{\mu}=k_{1}^{\mu}-q^{\mu}$. Double-copying with scalar QCD we find
\begin{align}
    \label{eq:3ptsGravVector}
    \mathcal{M}^{\text{H},s=1}_{3}&=\mathcal{A}^{s=0}_{3}\mathcal{A}^{\text{H},s=1}_{3},
\end{align}
where $\mathcal{M}^{\text{H},s=1}_{3}$ is the amplitude derived from \cref{eq:LagrGravitons1}
after applying the field redefinition in \cref{eq:GR1Redef}.

Compton scattering for spin-1 HQET is given by the kinematic numerators
\begin{subequations}
\begin{align}
	n_s^{\text{H},s=1} &= 2m\varepsilon_{v}^{*\beta}\varepsilon_{v}^{\alpha}\left[v\cdot\epsilon_{q_{1}}^{*}v\cdot\epsilon_{q_{2}}^{*}\eta_{\alpha\beta}\right.\notag \\
	&\ +\frac{v_{\rho}}{m}(\eta_{\alpha\nu}\eta_{\beta\mu}-\eta_{\alpha\mu}\eta_{\beta\nu})(\epsilon_{q_{1}}^{*\mu}q_{1}^{\nu}\epsilon_{q_{2}}^{*\rho}+\epsilon_{2}^{*\mu}q_{2}^{\nu}\epsilon_{q_{1}}^{*\rho})\notag \\
	&\ \left.-\frac{v\cdot q_{2}}{2m}(\epsilon_{q_{1}\alpha}^{*}\epsilon_{q_{2}\beta}^{*}-\epsilon_{q_{2}\alpha}^{*}\epsilon_{q_{1}\beta}^{*})\right] ,\\
	n_t^{\text{H},s=1} &= 0 ,\\
    n^{\text{H},s=1}_{u}&= n^{\text{H},s=1}_{s}|_{q_{1}\leftrightarrow q_{2}},
\end{align}
\end{subequations}
where, for brevity, we again write the numerators up to $\mathcal{O}(m^{-1})$ and in the case where $k_{1}=\epsilon_{i}\cdot\epsilon_{j}=q_{i}\cdot\epsilon_{j}=0$.
We have performed the calculation up to $\mathcal{O}(m^{-2})$ and checked the general case explicitly.
The double copy becomes
\begin{align}
    \label{eq:ComptGravVector}
    \mathcal{M}^{\text{H},s=1}_{4}&=\frac{n_{s}^{s=0}n_{s}^{\text{H},s=1}}{d_{s}}+\frac{n_{t}^{s=0}n_{t}^{\text{H},s=1}}{d_{t}}+\frac{n_{u}^{s=0}n_{u}^{\text{H},s=1}}{d_{u}},
\end{align}
where $\mathcal{M}^{\text{H},s=1}_{4}$ is derived from \cref{eq:LagrGravitons1} after applying the
field redefinition in \cref{eq:GR1Redef}.

Thus, we find that
\begin{align}
    (\text{QCD}_{s=0})\times(\text{HQET}_{s=1})=\text{HBET}_{s=1}
\end{align}
for three-point and Compton amplitudes.

\subsection{\texorpdfstring{$\frac{1}{2}\times\frac{1}{2}$}{1/2x1/2} Double Copy}

The spin-1 gravitational amplitudes can also be obtained by double-copying the spin-1/2 HQET amplitudes.
To do so, we use the on-shell heavy particle effective theory (HPET) variables of ref.~\cite{Aoude:2020onz} to modify eq.~(2.11) of ref.~\cite{Bautista:2019evw} for the case of heavy particles. Using the fact that the on-shell HPET variables correspond to momenta $p_{v}^{\mu}=m_{k}v^{\mu}$ with mass $m_{k}=m(1-k^{2}/4m^{2})$, following the derivation of ref.~\cite{Bautista:2019evw} leads to
\begin{align}
    \label{eq:1/2x1/2DC}
    \mathcal{M}^{\text{H},\frac{1}{2}\times\frac{1}{2}}_{n}&=\frac{m_{k_{1}}m_{k_{2}}}{m}\sum_{\alpha\beta}K_{\alpha\beta}\text{Tr}[\mathcal{A}^{\text{H},\frac{1}{2}}_{n,\alpha}P_{+}\slashed{\varepsilon}_{v}\bar{\mathcal{A}}_{n,\beta}^{\text{H},\frac{1}{2}}P_{-}\slashed{\varepsilon}_{v}^{*}],
\end{align}
where $P_{\pm}=(1\pm\slashed{v})/2$, $K_{\alpha\beta}$ is the KLT kernel, and $\alpha,\beta$ represent color orderings.
Here $\mathcal{A}^{\text{H}}$ and $\bar{\mathcal{A}}^{\text{H}}$ are amplitudes with the external states stripped,
and $\bar{\mathcal{A}}^{\text{H}}=-\gamma_{5}(\mathcal{A}^{\text{H}})^{\dagger}\gamma_{5}$.
We have also adopted the convention that only the initial matter momentum is incoming.
Converting to the on-shell HPET variables, it can be easily seen that 
\begin{align}
    \label{eq:PolVecDef}
    \varepsilon^{IJ}_{v\mu}(p)=\frac{1}{2\sqrt{2}m_{k}}\bar{u}^{I}_{v}(p)\gamma_{5}\gamma_{\mu}u^{J}_{v}(p),
\end{align}
with $I,J$ being massive little group indices.
Given the WNF for the heavy spinors, the WNF
for the polarization vector can easily be computed by comparing \cref{eq:PolVecDef} to its canonical polarization vector analog.
We find that it is indeed given by \cref{eq:InvR1}.

Applying \cref{eq:1/2x1/2DC} to \cref{eq:3ptsGluonSpinor} with the
three-point KLT kernel $K_{3}=1$,
we immediately recover the left-hand side of \cref{eq:3ptsGravVector}.
For Compton scattering the KLT kernel is
\begin{align}
    K_{4}=\frac{(s-m^{2})(u-m^{2})}{2q_{1}\cdot q_{2}}.
\end{align}
Then, applying \cref{eq:1/2x1/2DC} to the spin-1/2 HQET Compton amplitude with 
$k_{1},q_{i}\cdot\epsilon_{j},\epsilon_{i}\cdot\epsilon_{j}\neq0$ up to and including terms of order $\mathcal{O}(m^{-2})$,
we find \cref{eq:ComptGravVector} up to $\mathcal{O}(m^{-1})$.
When imposing $k_{1}=q_{i}\cdot\epsilon_{j}=\epsilon_{i}\cdot\epsilon_{j}=0$, cancellations make the double copy valid up to $\mathcal{O}(m^{-2})$.
The extension to higher inverse powers of the mass amounts
to simply including the contributions of higher-order operators in the HQET and HBET amplitudes.

Therefore, by using \cref{eq:1/2x1/2DC} to convert heavy spinors in amplitudes to heavy polarization vectors,
we have shown that
\begin{align}
    (\text{HQET}_{s=1/2})\times(\text{HQET}_{s=1/2})&=\text{HBET}_{s=1}
\end{align}
for three-point and Compton amplitudes.

\section{Conclusion}
\label{sec:Conclusion}

We have shown that the three-point and Compton amplitudes
derived from HQET can be double-copied to those of HBET for spins $s\leq1$.
As long as the matter states of HQET and HBET are related through the double copy,
in the sense described in \Cref{sec:CKRev},
and as long as higher-point amplitudes obey the spectral condition of ref.~\cite{Johnson:2020pny},
we see no obstacles to extending the double copy to higher-point amplitudes.

As mentioned in the introduction, due to the operator expansion of HPETs,
the double-copy relation between HQET and HBET can be studied at each order in the $\hbar$ expansion, with the 
classical limit being of special interest. 
%
%
Studying the double copy of HPETs through this lens
may provide some insight into the connection between the double copy with matter at the quantum and classical levels.
We leave this study for future work.

\acknowledgements
This project has received funding from the European Union's Horizon 2020 
research and innovation programme under the Marie Sk\l{}odowska-Curie grant 
agreement No. 764850 "SAGEX".
The work of A.H. was supported in part by the Danish National Research Foundation (DNRF91) and
the Carlsberg Foundation.

\begin{widetext}
\appendix

\section{Lagrangians for heavy particles}\label{sec:Lagrangians}

We present Lagrangians for heavy particles coupled to bi-adjoint scalars, gluons, and gravitons.
The heavy-particle Lagrangians were used to derive the scattering amplitudes in the paper.
For clarity, we omit the subscript $v$ for the heavy spin-1 fields.

\subsection*{Bi-adjoint scalars and heavy particles}

We couple the bi-adjoint scalars $\bf \Phi$ to heavy particles with spins $s\leq1$.
The spin-0 Lagrangian is
\begin{align}
	\label{eq:LagrBiadjoints0}
	\mathcal{L}_{\rm bi-adjoint}^{s=0} &= \phi_v^*\left[
		iv\cdot \partial - \frac{\partial_\perp^2 - y_s {\bf \Phi}}{2m} 
		+ \left( \frac{\partial_\perp^2 - y_s{\bf \Phi}}{2m}\right)
		\frac{1}{2m + iv\cdot \partial + \frac{\partial_\perp^2 - y_s{\bf\Phi}}{2m}}
		 \left( \frac{\partial_\perp^2 - y_s{\bf \Phi}}{2m}\right)
	\right] \phi_v.
\end{align}
The spin-1/2 Lagrangian is
\begin{align}
	\label{eq:LagrBiadjoints1/2}
	\mathcal{L}_{\rm bi-adjoint}^{s=1/2} &= \overline{Q}_v \left[
		iv\cdot \partial + y_f{\bf \Phi}
		+
		\left( i \slashed \partial_\perp\right)
		 \frac{1}{2m + iv\cdot \partial - y_f{\bf \Phi}}
		\left( i \slashed \partial_\perp\right)
	\right] Q_v.
\end{align}
The spin-1 Lagrangian is
\begin{subequations}
	\label{eq:LagrBiadjoints1}
\begin{align}
	\mathcal{L}_{\rm bi-adjoint}^{s=1} &= 
	- B_\mu^* (iv\cdot \partial) B^\mu  - \frac{1}{4m}B_{\mu\nu}^* B^{\mu\nu}
	+ \frac{y_v }{2m} B_\mu^* {\bf \Phi}B^\mu
	- \left( \mathcal{F}_-^\lambda B_\lambda^*\right)
	\frac{2}{m + \frac{1}{m}\partial_\perp^2}
	\left( \mathcal{F}_+^\lambda B_\lambda^*\right)
\end{align}
where
\begin{align}
	\mathcal{F}_\pm^\mu = \left( \pm \frac{i}{2}\partial^\mu - \frac{1}{2m} \partial^\mu (v\cdot \partial)
	+ \frac{y_v {\bf \Phi}}{2m}\right).
\end{align}
\end{subequations}
The coupling constants between the bi-adjoint scalars and the heavy scalars, fermions, and vectors are $y_s$, $y_f$, and $y_v$, respectively.

\subsection*{Gluons and heavy particles}
We couple gluons to heavy particles. 
The covariant derivative in this case is given by $D_{\mu}=\partial_{\mu}+ig_{s}\mathbf{T}^{a}A^{a}_{\mu}$.
The scalar Lagrangian is
\begin{align}
	\label{eq:LagrGluons0}
	\mathcal{L}_{\rm gluon}^{s=0}&=\phi_v^*\left[ iv\cdot D - \frac{D_\perp^2}{2m}
	+ \left(\frac{D_\perp^2}{2m}\right) \frac{1}{2m + iv\cdot D + \frac{D_\perp^2}{2m}} 
	\left( \frac{D_\perp^2}{2m}\right)\right]\phi_v.
\end{align}
The spin-1/2 Lagrangian is
\begin{align}
	\label{eq:LagrGluons1/2}
	&\mathcal{L}_{\rm gluon}^{s=1/2} = \overline{Q}_v \left[ iv\cdot D 
+ \left( i\slashed D_\perp\right) \frac{ 1}{2m+iv\cdot D }\left(i \slashed D_\perp\right)\right] Q_v .
\end{align}
The spin-1 Lagrangian \cite{Carone:1990pv} with gyromagnetic ratio $g=2$ can be written as
\begin{subequations}
    \label{eq:LagrGluons1}
\begin{align}
	\cL_{\rm gluon}^{s=1} =& - B_\mu^* (iv\cdot D) B^\mu - \frac{1}{4m}B_{\mu\nu}^* B^{\mu\nu} 
	+ \frac{ig}{2m} F^{\mu\nu}B_{\mu}^* B_\nu 
	- \left(\mathcal{E}^\lambda_- B_\lambda^*\right)
	\frac{2}{m + \frac{1}{m} D_\perp^2} \left(\mathcal{E}^\mu_+ B_\mu\right) 
\end{align}
where
\begin{align}
	\mathcal{E}_\pm^\mu =  \left(\pm \frac{i}{2}D^\mu - \frac{1}{2m} D^\mu(v\cdot D) \pm
	\frac{ig v_\nu F^{\nu\mu}}{2m}\right).
\end{align}
\end{subequations}
The heavy spin-1 states described by this Lagrangian are related to the canonical massive spin-1 states through
\begin{align}
	\label{eq:spin1effectiveFull}
    A^\mu(x) &=\frac{e^{-imv\cdot x}}{\sqrt{2m}} \left[ \delta^\mu_\nu - \frac{1}{1+iv\cdot \partial/m}\frac{iv^{\mu}\partial_{\nu}}{m}\right]B^\nu(x).
\end{align}
To obtain the desired heavy spin-1 states we apply the field redefinition
\begin{align}
	B_\mu \rightarrow \left[ \delta^\nu_\mu + \frac{1}{2m^{2}}\left(-v_{\mu}v\cdot D + D_\mu \right)D^\nu\right]B_\nu+\mathcal{O}(m^{-3}).
\end{align}

\subsection*{Gravitons and heavy particles} 
We couple gravitons to heavy particles.
The spin-0 Lagrangian is
\begin{subequations}
	\label{eq:LagrGravitons0}
\begin{align}
	&\sqrt{-g}\mathcal{L}_{\rm graviton}^{s=0} = 
	\sqrt{-g} \phi_v^* \left[ \mathcal{A}_1 + \left(\mathcal{A}_{2-}\right)\frac{1}{2m +i(v^\mu \nabla_\mu + \nabla_\mu v^\mu)- \mathcal{A}_1} \left(\mathcal{A}_{2+}\right)  \right] \phi_v,
\end{align}
where
\begin{align}
	\mathcal{A}_1 =&  \frac{1}{2}i g^{\mu\nu}(v_\mu \nabla_\nu + \nabla_\mu v_\nu) 
	+ \frac{1}{2}m(g^{\mu\nu}-\eta^{\mu\nu}) v_\mu v_\nu  
    - \frac{1}{2m} \nabla_\mu \left( (g^{\mu\nu}-\eta^{\mu\nu}) \nabla_\nu + \eta^{\mu\nu}\nabla_{\perp \nu}\right), \\
	\mathcal{A}_{2\pm} =& \frac{1}{2m} (im v_\mu - \nabla_\mu)\left( (g^{\mu\nu}-\eta^{\mu\nu})(-imv_\nu + \nabla_\nu)\right) 
    - \frac{1}{2m} \nabla_\mu (\eta^{\mu\nu}\nabla_{\perp\mu}) \pm \frac{1}{2}i \left[ \nabla_\mu v^\mu\right],
\end{align}
\end{subequations}
with $v^\mu \equiv \eta^{\mu\nu}v_\nu$ and $\nabla_{\perp\mu} \equiv \nabla_\mu - v_\mu (v^\nu \nabla_\nu)$.
%
%
The spin-1/2 Lagrangian is
\begin{subequations}
	\label{eq:LagrGravitons1/2}
\begin{align}
	\sqrt{-g}\mathcal{L}&_{\rm graviton}^{s=1/2}
	= \sqrt{-g}\,\, \overline{Q}_v\left[  i\slashed \nabla + \mathcal{B}
	 + \left(i\slashed \nabla + \mathcal{B} \right)
	P_-\frac{1}{2m -( i\slashed\nabla + \mathcal{B})P_-} \left(i\slashed \nabla + \mathcal{B}\right)\right] Q_v,
\end{align}
where $\slashed \nabla \equiv \delta^\mu_a \gamma^a \nabla_\mu$ and 
\begin{align}
	\mathcal{B} = (e^\mu_a - \delta^\mu_a)(i\gamma^a\nabla_\mu + m\gamma^a v_\mu).
\end{align}
\end{subequations}
The spin-1 Lagrangian can be written as
\begin{subequations}
    \label{eq:LagrGravitons1}
\begin{align}
	\sqrt{-g}\cL_{\rm graviton}^{s=1} =& \sqrt{-g} \left[
		-\frac{m}{2} (v_\mu B_\nu^*)(v_\rho B_\sigma)\left((g^{\mu\rho}-\eta^{\mu\rho})g^{\nu\sigma}
		-(g^{\mu\sigma}-\eta^{\mu\sigma})(g^{\nu\rho}-\eta^{\nu\rho})\right) \right. \nonumber \\
		& + \frac{i}{2}\left[\left(\nabla_\mu B_\nu^*\right)(v_\rho B_\sigma)
	-	\left(v_\mu B_\nu^*\right)(\nabla_\rho B_\sigma)
	\right]\left( g^{\mu\rho}g^{\nu\sigma}
		- g^{\mu\sigma}g^{\nu\rho}\right) \nonumber \\ 
		&- \left.\frac{1}{4m} B_{\mu\nu}^* B_{\rho\sigma}g^{\mu\rho}g^{\nu\sigma}
		- \left( \mathcal{C}_-^\alpha B_\alpha^*\right) \frac{1}{\mathcal{D}}
		\left( \mathcal{C}_+^\beta B_\beta\right)
	\right],
\end{align}
where
\begin{align}
	\mathcal{C}_\pm^\alpha =& -\frac{m}{2} (g^{\alpha\nu}-\eta^{\alpha\nu})v_\nu
	\pm \frac{i}{2}v_\nu \left[g^{\mu\rho} g^{\alpha\nu} 
	- g^{\alpha\mu}g^{\nu\rho}\right] \nabla_\mu 
	\left(v_\rho \pm \frac{i}{m} \nabla_\rho\right), \\
	\mathcal{D} =& \frac{m}{2} (v_\nu v_\sigma g^{\nu\sigma}) + \frac{1}{2m}v_\nu \left[g^{\mu\rho}g^{\nu\sigma}
- g^{\mu\sigma} g^{\rho\nu}\right] \nabla_\mu\nabla_\rho v_\sigma.
\end{align}
\end{subequations}
Note that though the velocity four-vector is constant its covariant derivative does not vanish because of the metric connection.
The heavy spin-1 states described by this Lagrangian are related to the canonical massive spin-1 states through \cref{eq:spin1effectiveFull}.
%
%
To obtain the desired heavy spin-1 states we apply the field redefinition
\begin{align}
    \label{eq:GR1Redef}
	B_\mu &\rightarrow \left[ \delta^\nu_\mu + \frac{1}{2m^{2}}\left(-g^{\alpha\beta}v_{\alpha}D_{\beta}v_{\mu} + D_\mu \right)g^{\nu\lambda}D_\lambda\right]B_\nu\notag \\
	&\quad+\mathcal{O}(m^{-3}).
\end{align}
Extending this redefinition to higher orders in $1/m$ is straight-forward.

\end{widetext}
\bibliography{double-copy-heavy-particles}

\end{document}